\documentclass{elsart}

\usepackage{multicol}
\usepackage{amssymb}

\begin{document}

\begin{frontmatter}



\title{New Trends in the Zee Model}


\author{Yoshio Koide}
\ead{koide@u-shizuoka-ken.ac.jp}
\address{ Department of Physics, University of Shizuoka \\
          52-1 Yada,\ Shizuoka 422-8526,\ Japan}

\begin{abstract}
Recent trends in the Zee model are reviewed.
Especially, the importance of a serious constraint in the Zee 
model, $\sin^2 2\theta_{solar} =1.0$, is pointed out.
\end{abstract}

\end{frontmatter}

\begin{multicols}{2}
\section{introduction}
\label{Sec1}

The Zee model \cite{Zee} is very attractive to us. 
The model has only three parameters and it can naturally  
give a large neutrino mixing.
The Zee mass matrix is given by the form
$$
M_\nu = m_0
\left( \begin{array}{ccc}
0 & a & c \\
a & 0 & b \\
c & b & 0
\end{array} \right) ,
\eqno(1.1)
$$
where
$$
\begin{array}{l}
a=f_{e \mu} (m_\mu^2 - m_e^2)  , \\ 
b=f_{\mu \tau} (m_\tau^2 - m_\mu^2) , \\ 
c=f_{\tau e} (m_e^2 - m_\tau^2) ,\\
\end{array}
\eqno(1.2)
$$
and $f_{ij}$ are lepton flavor violating Yukawa coupling constants
of the Zee scalar.
Especially, for the case 
$a=c\gg b$,
it leads to a bi-maximal mixing \cite{Jarlskog}
$$
U  \simeq 
\left( \begin{array}{ccc}
\frac{1}{\sqrt{2}} & -\frac{1}{\sqrt{2}} & 0 \\
\frac{1}{2} & \frac{1}{2} & -\frac{1}{\sqrt{2}} \\
\frac{1}{2} & \frac{1}{2} & \frac{1}{\sqrt{2}}
\end{array} \right) ,
\eqno(1.3)
$$
with $\Delta m_{12}^2/\Delta m_{23}^2
\simeq \sqrt{2} {b}/{a}$.
Besides the model can provide rich new physics,
especially, in lepton flavor violating processes.

\section{Present and future in the Zee model}
\label{Sec2}

{\it Can the model explain the observed $\Delta a_\mu$ ?}
The Zee model can provide us rich phenomenology:
radiative decays of the charged leptons $e_i^- 
\rightarrow e_j^- \gamma$,
two gamma decay of the CP even neutral Higgs 
scalar $h^0 \rightarrow \gamma \gamma$,
lepton flavor changing Z decays $Z \rightarrow  e_i^\pm e_j^\mp$,
and so on. 
(For example, see the references in Ref. \cite{G-K-F}.)
Especially, we are interested in the observed excess of 
muon anomalous magnetic moment $\Delta{a}_{\mu}$. Here,
We give a short review for typical two models.
One is a case with $|f_{e\mu}| \gg |f_{\tau e}| \gg 
|f_{\mu \tau}|$ proposed by Jarlskog {\it et al.}
\cite{Jarlskog}, and another is a case with
$|f_{\mu\tau}| \gg |f_{e\mu}| \gg |f_{\tau e}|$ proposed 
by Smirnov and Tanimoto \cite{Smirnov-Tanimoto},
The former leads to a nearly bimaximal mixing (1.3), so that we can
explain both the solar data, but we cannot explain the value of 
$\Delta{a}_{\mu}$ because of 
$|f_{\mu \tau}| \ll |f_{\tau e}| \ll |f_{\mu \tau}|$.
On the other hand, the latter can give the value of $\Delta{a}_{\mu}$
because of
$|f_{\mu\tau}| \gg |f_{e\mu}| \gg |f_{\tau\mu}|$,
but it leads to $\sin^2 2\theta_{12}=0$ and
$\sin^2 2\theta_{23}=1$,
and cannot explain the solar neutrino data.
In order to explain the solar neutrino data,
we must consider, for example, a sterile neutrino ${\nu}_{s}$
which mixes with ${\nu}_{e}$ \cite{Smirnov-Tanimoto},

{\it Can the model be related to the charged lepton masses ?}
The Zee coupling constants ${f}_{ij}$ are free parameters which
are irrelevant to Yukawa coupling constants ${y}_{i}^{(f)}$
($f=e,u,d$),
we must seek for a further ansatz for ${f}_{ij}$ in order to
relate ${f}_{ij}$ with the charged lepton masses and so on.
One of such attempts has been proposed by Koide 
and Ghosal \cite{K-G}
They have put a simple ansatz on the transition matrix elements
in the infinite momentum frame (not on the mass matrix),
and they have obtained the relations
$$
{f}_{ij}={\varepsilon}_{ijk} [{m_k^e}/(m_i^e+m_j^e)] f ,
\eqno(2.2)
$$
where $m_{i}^{e}=(m_e,m_\mu,m_\tau)$,
which leads to  the prediction
$$
R \equiv \frac{\Delta{m}_{12}^{2}}{\Delta{m}_{23}^{2}}
\simeq\sqrt2\frac{m_{e}}{m_{\mu}}
= 6.9 \times 10^{-3} .
\eqno(2.3)
$$
The predicted value $(2.3)$ is in excellent agreement with 
the observed value (best fit values) \cite{atm,Garcia}
$$
R_{exp} \simeq \frac{2.2 \times 10^{-5} {\rm eV}^2}{
3.2 \times 10^{-3} {\rm eV}^2}=6.9 \times {10}^{-3}.
\eqno(2.4)
$$
However, the meaning of the ansatz for matrix elements on the infinite
momentum frame is still unclear. Further study will be required.

{\it What experimental value is serious for the Zee model ?}
Recently, it has been pointed out that the Zee model cannot give 
the observed sizable deviation from $\sin^2{2\theta}_{solar} = 1$,
i.e., $\sin^2{2\theta}_{observ}^{solar}\sim 0.8$ under the condition 
$\Delta{m}_{solar}^2 / \Delta{m}_{atm}^2 \ll 1$: 
A parameter independent investigation leads to a severe constraint \cite{koide}
on the value of $\sin^2{2\theta}_{solar}$
$$
\sin^2{2\theta}_{solar} \geq {1}- \frac{1}{16} 
\left(\frac{\Delta m_{solar}^2}{\Delta m_{atm}^2}\right)^2.
\eqno(2.4)
$$
The conclusion cannot be loosened even if we take RGE effect into
consideration.

{\it How can the model be embedded into a GUT scenario ?}
Another problem in the Zee model is that the original Zee model
is not on a GUT scenario.
Where is a room of the Zee scalar $h^+$ in a GUT scenario?
The scalar $h^{+}$ belongs to
$(1.1)_{Y = 2}$ of SU(3)$_c \times$SU(2)$_L \times$U(1)$_{Y}$.
The candidates are as follows:
(i) slepton $\tilde{e}_{r}$, i.e., a member of $10_{\tilde{f}}$ of SU(5)
in a R-parity breaking SUSY model;
(ii) a member of messengers $10_{M} + \overline{10}_{M}$ of 
SUSY-breaking in a R-parity conserving SUSY model;
(iii)  a member of new hypothetical 10-plet scalar of SU(5).
These extensions can bring new additional contributions into the neutrino
masses.
Therefore, we will be free from the severe constraint
$\sin^2{2\theta}_{solar} = 1.0$.

\section{Summary}
\label{Sec3}

 The Zee model is very attractive to us, because the model
can provide us rich phenomenology.
The effort to relate the Zee coupling constants with the Yukawa coupling 
constants of the charged leptons will be important. An attempt has
been reviewed:
$f_{ij} = \varepsilon_{ijk}[m_{k}^e/(m_{i}^{e} + m_{j}^{e})]f$.
On the other hand, recently,
a serious constraint in the Zee model
has been reported. It is very interesting whether the experiments
rule out the value $\sin^2{2\theta}_{solar} = 1.0$ or not.
Attempts to embed the Zee model into a GUT scenario will become
more important in order to be free from the severe constraint
$\sin^2{2\theta}_{solar} = 1.0 $ on the original Zee model.


\end{multicols}

\begin{thebibliography}{99}

%
%
\bibitem{Zee} A.~Zee, Phys.~Lett. {\bf 93B}, 389 (1980); 
{\bf 161B}, 141 (1985); L.~Wolfenstein, Nucl.~Phys. {\bf B175},
93 (1980);
S.~T.~Petcov,  Phys.~Lett. {\bf 115B}, 401 (1982).
%
\bibitem{Jarlskog}
C.~Jarlskog, M.~Matsuda, S.~Skadhauge, M.~Tanimoto, 
Phys.~Lett. {\bf B449}, 240 (1999).%
%
%
\bibitem{G-K-F}  A.~Ghosal, Y.~Koide and H.~Fusaoka, hep-ph/0104104,
to be published in Phys.~Rev. {\bf D}.
%
\bibitem{Smirnov-Tanimoto}
A.~Yu.~Smirnov and M.~Tanimoto, Phys.~Rev. {\bf D55}, 1665 (1997).
%
\bibitem{K-G} Y.~Koide and A.~Ghosal, Phys.~Rev. 
{\bf D63}, 037301 (2001).
%
\bibitem{atm} Y.~Fukuda {\it et al.}, Phys.~Lett. {\bf B335}, 
237 (1994);
Super-Kamiokande collaboration, Y.~Fukuda, {\it et. al.},
Phys.~Rev.~Lett. {\bf 81}, 1562 (1998);
H.~Sobel, Talk presented at {\it Neutrino 2000},
Sudbury, Canada, June 2000 (http://nu2000.sno.laurentian.ca/).
%
\bibitem{Garcia} M.~Gonzalez-Garcia, Talk presented at {\it Neutrino 2000},
Sudbury, Canada, June 2000 (http://nu2000.sno.laurentian.ca/).
\bibitem{koide} Y.~Koide, hep-ph/0104226, to be published 
in Phys.~Rev. {\bf D}.
%
%
%
%
%
%
%
%
%
%
%
\end{thebibliography}
\end{document}